\documentclass[twocolumn,twocolappendix]{aastex6}

\begin{document}

\newcommand{\kms}{\ensuremath{\mathrm{km}\,\mathrm{s}^{-1}}}
\newcommand{\galunits}{\ensuremath{\mathrm{km}\,\mathrm{s}^{-1}\,\mathrm{kpc}^{-1}}}
\newcommand{\galacc}{\ensuremath{\mathrm{km}^2\,\mathrm{s}^{-2}\,\mathrm{kpc}^{-1}}}
\newcommand{\MLsun}{\ensuremath{\mathrm{M}_{\sun}/\mathrm{L}_{\sun}}}
\newcommand{\Lsun}{\ensuremath{\mathrm{L}_{\sun}}}
\newcommand{\Msun}{\ensuremath{\mathrm{M}_{\sun}}}
\newcommand{\Ha}{H$\alpha$}
\newcommand{\sfr}{\ensuremath{\psi_*}}
\newcommand{\avesfr}{\ensuremath{\langle \psi_* \rangle}}
\newcommand{\sfrate}{\ensuremath{\mathrm{M}_{\sun}\,\mathrm{yr}^{-1}}}
\newcommand{\Aunits}{\ensuremath{\mathrm{M}_{\sun}\,\mathrm{km}^{-4}\,\mathrm{s}^{4}}}
\newcommand{\surfdens}{\ensuremath{\mathrm{M}_{\sun}\,\mathrm{pc}^{-2}}}
\newcommand{\voldens}{\ensuremath{\mathrm{M}_{\sun}\,\mathrm{pc}^{-3}}}
\newcommand{\gevcc}{\ensuremath{\mathrm{GeV}\,\mathrm{cm}^{-3}}}
\newcommand{\etal}{et al.}
\newcommand{\LCDM}{$\Lambda$CDM}
\newcommand{\ML}{\ensuremath{\Upsilon_*}}
\newcommand{\Mst}{\ensuremath{M_*}}
\newcommand{\Mg}{\ensuremath{M_g}}
\newcommand{\Mb}{\ensuremath{M_b}}


\title{MOND Prediction for the Velocity Dispersion of the `Feeble Giant' Crater II}

\author{Stacy S. McGaugh\altaffilmark{1}}
\altaffiltext{1}{Department of Astronomy, Case Western Reserve University, Cleveland, OH 44106}

\begin{abstract}
Crater II is an unusual object among the dwarf satellite galaxies of the Local Group in that it has a very large size
for its small luminosity. This provides a strong test of MOND, as Crater II should be in the deep MOND regime
($g_{in} \approx 34\;\galacc \ll a_0 = 3700\;\galacc$). Despite its great distance ($\approx 120$ kpc) from the Milky Way,
the external field of the host ($g_{ex} \approx 282\; \galacc$) comfortably exceeds the internal field. 
Consequently, Crater II should be subject to the external field effect, a feature unique to MOND.
This leads to the prediction of a very low velocity dispersion: $\sigma_{efe} = 2.1^{+0.9}_{-0.6}\;\kms$.

\end{abstract}



\section{Introduction}
\label{sec:Intro}

The Modified Newtonian Dynamics (MOND) \citep{milgrom83} hypothesizes a modification of dynamics as an alternative
to non-baryonic dark matter. While {not a complete, relativistic theory
intended as a replacement for the standard model of cosmology} \citep{milgrombicosmo,CJP}, 
MOND makes very specific predictions for the dynamics of low acceleration systems \citep{milgromscholarpedia}. 
A remarkable number of these predictions have been realized \citep{SMmond,FM12}. 
The predictive ability of MOND is unexpected in the conventional
cold dark matter paradigm which, at present, provides no satisfactory explanation for it \citep{sandersNOCDM}.

In the absence of dark matter, low surface brightness galaxies are necessarily in the low acceleration regime
as a direct consequence of the diffuse nature of the luminous mass distribution. 
Indeed, one of the original predictions of MOND was that low surface brightness 
galaxies should exhibit large mass discrepancies \citep{milgrom83}.
Consequently, strong tests are provided by both late type, rotating LSB galaxies \citep{MdB98b,swatersmond} and
diffuse pressure supported systems like dwarf spheroidals \citep{GS92,milg7dw,angusdw,serradw}.
In the context of MOND, these systems appear to be dark matter dominated\footnote{This also applies
to Dragonfly 44 \citep{dragonfly44} and any other diffuse extragalactic system.}
because they are in the deep MOND regime, $g_{in} \ll a_0$.
Numerically, $a_0 = 1.2 \times 10^{-10}\;\mathrm{m}\,\mathrm{s}^{-2} = 3700\;\galacc$ \citep{BBS,M11}.

At this juncture, it is clear that the dynamical surface density scales with surface brightness \citep{SPARCIII}.
{This was not anticipated by conventional galaxy formation simulations \citep{OmanDiversity}, but is
expected in} MOND \citep{milgrom2016}. While the success of MOND in fitting rotation curves is widely known \citep{SMmond},
it has also had considerable success in predicting the velocity dispersions of the dwarf satellites of 
Andromeda \citep{MM13a,MM13b,PM2014}. Many or these predictions were made in 
advance\footnote{Ten cases were predicted in advance by \citet{MM13a}, three by \citet{MM13b}, and 
three more by \citet{PM2014}. See also \citet{PMJ2015} for further (as yet mostly untested) predictions.}
of the measurement of the velocity dispersion, so constitute true \textit{a priori} predictions.

Each and every galaxy poses a distinct test.
New objects with extreme properties provide a good opportunity to test the theory
by pressing it into a regime where it is not already well tested.
The case of the `feeble giant' Crater II \citep{CraterII} provides such an opportunity.

Crater II is a remarkable object, having a huge linear size (half light radius $r_h \approx 1.1$ kpc)
for its tiny luminosity ($M_V \approx -8$) \citep{CraterII}. This is like a single globular cluster stretched
out to be the size of a galaxy. The only comparable object currently known is And XIX \citep{Collins2014}.

Given its extremely low surface brightness, Crater II should be deep in the MOND regime ($g_{in} \ll a_0$). 
It is so feeble that despite its great distance from the Milky Way, it may be subject to the External Field Effect (EFE).  
The EFE is a strange feature of MOND in which the dynamics of a dwarf satellite can be affected by the field of its host.
Whether the EFE affects Crater II is an important test, as the EFE
is unique to MOND and cannot plausibly be attributed to baryonic effects in \LCDM.

\section{Data and Context}
\label{sec:data}

We adopt for Crater II the values published by \citet{CraterII}:
a half-light radius $r_h = 1066 \pm 84$ pc and an absolute magnitude $M_V = -8.2 \pm 0.1$.
The latter corresponds to a luminosity $L_V = 1.6 \times 10^5\;\Lsun$.
Translating the heliocentric distance to a Galactic coordinate frame, Crater II is about 120 kpc
from the Galactic center.

Given its large size, the stars have the opportunity to probe a large range of the gravitational potential well.
If Crater II resides in a dark matter sub-halo, one would thus expect a high velocity dispersion
for this type of galaxy. The empirical scaling relation between size and velocity dispersion \citep{walkerandme}
anticipates $\sigma \approx 17.5\;\kms$. There is large scatter about this relation, so a wide range of
velocity dispersions might seem plausible in \LCDM, which makes no specific prediction for each individual
dwarf. {Nevertheless, the large size of Crater II anticipates a relatively large velocity dispersion
if it resides in a dark matter halo with an NFW-like potential.}

Indeed, it came as a surprise that comparably large dwarfs could have rather low velocity 
dispersions \citep{Collins2014}.  In particular, Crater II is similar to And XIX in having a large size for its
luminosity. And XIX and several other dwarfs of Andromeda have low velocity dispersions that were
correctly predicted by MOND \citep{MM13a,MM13b}. Consequently, it is interesting to make the
prediction specifically for Crater II.

\section{MOND Prediction}
\label{sec:pred}

We predict the velocity dispersion of Crater II using the same methods
applied previously to other dwarf satellite galaxies \citep{MM13a}. 
We assume isotropic orbits in a spherical system, and estimate the 3D
half-mass radius as $r_{1/2} = 4 r_h/3$ \citep{boom}.
We adopt a stellar mass-to-light ratio of  
$\ML = 2^{+2}_{-1}\;\MLsun$. The uncertainty in this mass-to-light ratio
dominates the formal sources of observational error, which we will ignore
bearing in mind the considerable uncertainty in these uncertainties.

The velocity dispersion of an isolated
spherical galaxy of mass $M_*$ in the deep MOND limit 
is \citep{milg7dw,MM13a}
\begin{equation}
\sigma_{iso} \approx \left( \frac{4}{81} a_0 G M_* \right)^{1/4}.
\label{eq:isolatedMOND}
\end{equation}
Numerically, this becomes $(\sigma_{iso}/\kms) = (M_*/1264\;\Msun)^{1/4}$.
For Crater II, $M_* \approx 3.3 \times 10^5\;\Msun$, so $\sigma_{iso} \approx 4\;\kms$.

This value for the velocity dispersion of $4\;\kms$ is what MOND would predict for
a diffuse, isolated galaxy of this stellar mass. This is rather less than one would naturally 
anticipate for a typical sub-halo \citep{2B2F,GKBKBK14}, and certainly 
less than expected from the scaling relation of \citet{walkerandme}.
However, MOND predicts the velocity dispersion to be smaller still if
this dwarf is subject to the external field effect (EFE).

The EFE  \citep{BM84,milgrom2014} is a unique feature of MOND. 
Whether it is present or absent thereby provides a unique test above and beyond
tests in isolated systems (e.g., rotation curves). The EFE {is a consequence of the
nonlinearity\footnote{MOND violates the strong equivalence principle \citep{Will2014LRR}, 
but not the universality of free fall \citep{Milgrom2015}.} of MOND: the acceleration of a star is not
simply the vector sum $\sum_i \vec{g}_{*,i}$ 
of all other stars acting separately on it as it is in linear theories like Newtonian 
gravity. It} arises when a very low acceleration system (like a dwarf satellite
with internal field $g_{in}$) is embedded in the field of a merely low acceleration system 
(like a host galaxy with external field $g_{ex}$) such that $g_{in} < g_{ex} < a_0$ \citep{milg7dw}.
{The imposition of an external field alters the dynamics of a feeble system because it is
not as deep in the MOND regime as it would be if isolated.}

In the EFE regime, the velocity dispersion estimator of MOND becomes \citep{MM13a} 
\begin{equation}
\sigma_{efe} \approx \left(\frac{a_0 G M_*}{3 g_{ex} r_{1/2}}\right)^{1/2}.
\label{eq:MONDEFE}
\end{equation}
This has the same structure as the Newtonian estimator but with an effective
value of Newton's constant enhanced by the MOND interpolation function
$G_{eff} = G/\mu(g/a_0)$. In the case of Crater II, 
$\mu(g/a_0) \approx g_{ex}/a_0$ is an excellent approximation.

For an isolated system, the MOND-predicted velocity dispersion depends only on 
mass (eq.~\ref{eq:isolatedMOND}).  In the EFE regime, it additionally depends on
the size of the system and the external field.  
Consequently, two dwarfs that are photometrically identical
should\footnote{This should not happen in linear theories like Newtonian gravity,
with or without dark matter.} have different velocity dispersions if one is 
isolated while the other is subject to the EFE. \citet{MM13b} identify 
several such pairs of dwarfs around Andromeda 
where this predicted difference appears to be reflected in the data.

The exceptional size of Crater II makes it likely to be subject to the EFE
despite its great distance from its host. We estimate the internal
acceleration at the half-light radius \citep{MM13a} as
\begin{equation}
g_{in} \approx \frac{3 \sigma_{iso}^2}{r_{1/2}}
\label{eq:intacc}
\end{equation}
where $\sigma_{iso}$ is calculated with eq.~\ref{eq:isolatedMOND}.
For Crater II, $g_{in} \approx 34\;\mathrm{km}^2\;\mathrm{s}^{-2}\;\mathrm{kpc}^{-1} = 0.009 a_0$.
This is lower than all of the many dwarfs considered by \citet{MM13a} except for And XIX, 
to which it is equal.

The external acceleration we estimate as
\begin{equation}
g_{ex} \approx \frac{V_{MW}^2}{D_{GC}},
\label{eq:extacc}
\end{equation}
where $D_{GC} = 120$ kpc is the Galactocentric distance and the circular velocity of
the Milky Way at this distance is taken to be $V_{MW} = 184\;\kms$ \citep{M2016}. 
This works out to $g_{ex} \approx 282\;\mathrm{km}^2\;\mathrm{s}^{-2}\;\mathrm{kpc}^{-1} = 0.076 a_0$,
an order of magnitude larger than the internal acceleration of Crater II.
Consequently, Crater II is in the EFE regime. 

The predicted velocity dispersion of Crater II is thus
\begin{equation}
\sigma_{efe} = 2.1^{+0.9}_{-0.6}\;\kms. 
\label{eq:pred}
\end{equation}
The uncertainty here represents a factor of two range around the assumed 
mass-to-light ratio. Observational errors are neglected.

This velocity dispersion is smaller than the isolated MOND case by a factor of two,
and nearly an order of magnitude smaller than anticipated by scaling 
relations \citep{walker,walkerandme}. MOND predicts that Crater II should be discrepant
from these scaling relations in the same sense as And XIX \citep{Collins2014}.
The low velocity dispersion of And XIX was correctly predicted by
MOND \citep{MM13a,MM13b} for the same reason: the external field dominates.

\section{Discussion}
\label{sec:disc}

\subsection{Potential Complications}

Our predictions are only as good as the input data and the assumptions that underly them. 
We assume spherical symmetry, isotropic orbits, and dynamical equilibrium.
Deviations from these ideals could affect the predicted velocity dispersion.

At the level of a few \kms, many potential systematics (like interlopers)
can affect the measured velocity dispersion. Essentially all act to artificially inflate $\sigma$
\citep[see discussion in][]{MWolf}. Indeed, the motions of binary 
stars can by themselves contribute at the $\sim 2\;\kms$ 
level \citep[e.g.,][]{MC2010,Simon2011,WalkerRet2}.

Despite its great distance from the Galactic center,
Crater II is so diffuse that it may be subject to tidal effects.
Its tidal radius in MOND \citep{zhaotian} is about $\sim 1.6$ kpc, not vastly larger 
than its half-light radius. This is an uncertain calculation specific
to its current orbital distance, so must be interpreted with caution.
However, should evidence of tidal disruption emerge, 
it would be more natural in MOND than if Crater II inhabits a 
dark matter sub-halo, as the latter acts to shield stars from tides.

As a dwarf orbits its host, the external field varies,
depending on the eccentricity and orientation of the orbit.
\citet{bradadwarf} introduced a parameter $\gamma$ as a measure of the 
severity of temporal variation in the external field (see their equation 7). 
It can be interpreted as the typical number of internal orbits a
star makes in the time it takes the satellite galaxy to complete
one orbit around the host. For large $\gamma$, stars complete many
internal orbits for every orbit of the dwarf around the host, so the dwarf
can adapt to changes in the external field adiabatically. 
For small $\gamma$, dwarfs may not have time to adjust.

Many ultrafaint dwarfs show hints of tidal effects at low values
of $\gamma \lesssim 8$ \citep{MWolf}. This suggests
a violation of the assumption of dynamical equilibrium.
For Crater II, $\gamma \approx 2$ for a star at the half-light radius, 
so the assumption of equilibrium is a concern:
such a star orbits the Milky Way half as often as it orbits within the dwarf.  

There is reason to hope that Crater II may nevertheless provide a robust test. 
Unlike the ultrafaint dwarfs discussed by \citet{MWolf}, it is sufficiently
far out that the time scales are long. The orbital period at its
current distance from the Milky Way is $\sim 4$ Gyr.
Thus it has likely made only a few orbits in a Hubble time, 
and any non-equilibrium effects may as yet be subtle.

\subsection{Nominal Expectation in \LCDM}
\label{sec:LCDM}

Unlike MOND, \LCDM\ does not make a specific prediction for the 
velocity dispersions of individual dwarfs. 
As a population, one expects them to fall along the extrapolation
of the stellar mass--halo mass relation \citep[e.g.,][]{Mosterreln,Behroozi}.
Extrapolating from abundance matching to these low mass scales is very uncertain.
To complicate matters further, there must be a great
deal of scatter in mapping from luminosity to halo mass
simply from halo to halo scatter \citep{TBGW2011}.  
This can be seen in the simulations of \citet{BZ2014}, in which
the characteristic velocity varies by a factor of $\sim 3$ for galaxies
of similar luminosity (see their Fig.~8, which depicts circular speeds at
the scale of 1 kpc appropriate to Crater II).
Consequently, there is no agreed method by which the observed
distribution of the stars can be used to predict the velocity dispersions
of dwarfs in \LCDM. Given the expected scatter, 
it is not obvious that such an exercise should
even be possible, despite the success of \citet{MM13a} in doing it.

To get a sense of the expected scale of the velocity dispersion,
we first consider pure NFW halos \citep{NFW} of the appropriate
mass scale. Because of the highly non-linear relation between 
luminosity and mass required by abundance 
matching \citep{Mosterreln,Behroozi}, low luminosity dwarfs
are currently expected to reside in halo masses 
of $\lesssim 10^{10}\;\Msun$ \citep[e.g.,][]{BDC2015,DuttonSimV,coresallthewaydown}.
At the half light radius of Crater II, an NFW halo with 
$M_{200} = 10^{10}\;\Msun$ has a circular velocity of $29\;\kms$,
corresponding to $\sigma \approx 17\;\kms$. This corresponds well
to the size-velocity dispersion scaling relation of \citet{walkerandme}.

Fig.~1 of \cite{2B2F} provides a good illustration of how small velocities
measured at large radii are problematic for \LCDM. In order to obtain
a velocity dispersion as low as that predicted by MOND, we need to 
consider smaller NFW halos. Knowing the circular velocity curve of
an NFW halo, this is trivial to compute for a specified 
cosmology \citep{cV200reln}.  At $M_{200} = 10^{9}\;\Msun$, 
$\sigma \approx 10\;\kms$ at $r \approx 1$ kpc. At 
$M_{200} = 10^{8}\;\Msun$, $\sigma \approx 6\;\kms$.
We do not reach comparably low velocity dispersions until
the absurdly small halo mass $M_{200} = 10^{7}\;\Msun$
for which $\sigma \approx 3\;\kms$.  At this scale, the rotation
curve peaks and begins to decline well within the half-light radius,
and the extent of the halo is not great ($R_{200} \approx 4.5$ kpc).

So far we have only considered `raw' NFW halos.
There is a rich field of work on how such halos may be modified by baryonic 
effects \citep[e.g.,][]{BZ2014,FIRE2015,BDC2015,DuttonSimV,coresallthewaydown}.
It is far beyond the scope of this letter to review this topic, which has many different
(and often divergent) implementations of feedback and other sub-grid physics.
For our purposes, it suffices to note that it may be possible to ascribe a
factor of $\sim 2$ reduction in velocity dispersion to baryonic 
effects \citep{coresallthewaydown}, so perhaps we might 
anticipate $\sigma \approx 8\;\kms$ rather than $17\;\kms$.
Either way, the expectation of \LCDM\ is for a higher 
velocity dispersion than that predicted by MOND.

\subsection{Prediction and Accommodation}

It would be good to have a test that clearly distinguished 
between \LCDM\ and MOND. Some observations \citep{SMmond,FM12} 
prefer a dark matter interpretation (e.g., clusters of galaxies) while
others prefer MOND (e.g, rotation curves). Which interpretation
seems preferable depends on how we weigh the various lines
of evidence \citep{CJP}. In general, the dark matter paradigm can
accommodate a broader range of phenomena, while MOND has
had {many predictive success that are unexpected in the context
of} \LCDM\ \citep{SMmond,FM12,CJP}.

An important aspect to consider in weighing the evidence is the uniqueness
of the prediction each theory makes. MOND makes a very specific
prediction for rotation curves that is unique to each individual 
galaxy \citep{sandersNOCDM}.
\LCDM\ does not. It therefore seems strange that out of the enormous
parameter space available to galaxies constructed of a baryonic disk plus
dark matter halo that the result so often looks like MOND \citep{SPARCIV}.
This is a fine-tuning problem \citep{CJP}.

Objects like Crater II provide an additional uniqueness test.
Conventionally, if Crater II has a velocity dispersion of $\approx 0.6\;\kms$
we would say it is a star cluster with no dark matter. If it has a velocity
dispersion of $4\;\kms$, we would say it has some dark matter.
If it has a velocity dispersion of $8\;\kms$, we would say it has lots of dark matter.
If $\sigma \approx 17\;\kms$, we would declare it to be one of the most dark matter
dominated galaxies known.
There is no uniquely predicted vale; we simply infer the amount of dark matter that
corresponds to the observed velocity dispersion.
\LCDM\ can accommodate any result.

In MOND, the prediction is unique.
Barring a drastic failure of the input data and necessary assumptions, 
the velocity dispersion of Crater II must be $\sigma_{efe} \approx 2\;\kms$.
It should not be $4\;\kms$ as it would be if Crater II were isolated.
It certainly should not be significantly larger: 
a high observed velocity dispersion would falsify MOND.

In contrast, a velocity dispersion as low as $2\;\kms$ would be unprecedented 
in \LCDM. The sub-halos in which dwarfs of comparable size and luminosity 
are imagined to reside are expected to have higher velocity dispersions \citep{2B2F},
especially for objects extending over kpc scales. 
The small amplitude and uniqueness of the value predicted by MOND would thus 
pose a problem for \LCDM, were it to be observed. 

\begin{acknowledgements}
This publication was made possible through the support of the John Templeton Foundation. 
The opinions expressed here are those of the authors and do not necessary reflect the views of the John Templeton Foundation.
\end{acknowledgements}

\end{document}